\documentclass[prl,twocolumn,showpacs,amsmath,amssymb]{revtex4}

\usepackage{graphicx}
\usepackage{dcolumn}
\usepackage{bm}

\include{revmacro}

\begin{document}


\title{Vector chiral spin liquid phase in absence of geometrical frustration \\
}


\author{Fabio~Cinti$^{(1,2)}$}
\email{cinti@ualberta.ca}
\author{Alessandro~Cuccoli$^{(2)}$}
\author{Angelo~Rettori$^{(2)}$}
\affiliation{$^{(1)}$Department of Physics, University of Alberta, Edmonton, Alberta, Canada T6G 2J1}
\affiliation{$^{(2)}$Department of Physics, University of Florence and CNISM, 50019 Sesto Fiorentino (FI), Italy.}

\date{\today}

\begin{abstract}

Making use of detailed classical Monte Carlo simulations, we study the critical properties of a  two dimensional planar spin model on a square lattice composed by weakly interacting helimagnetic chains. We find a large temperature window where the vector chirality order parameter,  $\langle{{\bm \kappa}_{jk}}\rangle$=$\langle{\bf{S}}_j\times{\bf{S}}_k\rangle$, the key  quantity in multiferroic systems, takes nonzero value in absence of long-range order or quasi-long-range order, so that, our model is the first example where, at finite temperatures, a vector chiral spin liquid phase in absence of geometrical frustration is explicitly reported. We also show that the strength of interchain interaction is fundamental in order to obtain the vector chiral spin liquid phase.
The relevance of our results for three-dimensional models is also discussed.
\end{abstract}

                         
\maketitle

Geometrical frustration and/or competition between interactions can lead to exotic noncollinear 
magnetic thermodynamic phases, which can be characterized by unusual order parameters. Particularly 
relevant are two order parameters: scalar chirality $\langle\chi_{jkl}\rangle$=$\langle{\bf{S}}_j\times{\bf{S}}_k 
\cdot{\bf{S}}_l\rangle$ and vector chirality (or spin current)  $\langle{{\bm 
\kappa}_{jk}}\rangle$=$\langle{\bf{S}}_j\times{\bf{S}}_k\rangle$. These two chiralities present different symmetries: 
nonzero value of  $\langle\chi_{jkl}\rangle$ implies that the time-reversal symmetry is broken, while parity 
symmetry breaking comes with $\langle{\bm{\kappa}}_{jk}\rangle$$\not=$0. Both of them are relevant in strongly
correlated electron systems: a nonzero $\langle\chi_{jkl}\rangle$ gives rise to large anomalous Hall effect 
\cite{Machida2007} and leads to orbital electric currents in frustrated geometries 
\cite{Bulaevskii2008}, while new electromagnetic phenomena emerge in Mott insulators as a 
consequence of induced $\langle\chi_{jkl}\rangle$, generated by the coupling between the  $\langle{{\bm 
\kappa}_{jk}}\rangle$
and an external homogeneous magnetic field \cite {Al-Hassanieh2009}. On the other hand, 
relativistic spin-orbit interaction leads to a coupling between the vector chirality and the 
electric polarization \cite{Katsura2005,Mostovoy2006,Sergienko2006,Cheong2007} which play a fundamental role 
in magnetoelectric properties. This coupling permits also to obtain experimental informations about 
the vector chirality (which is difficult to measure owing to the absence of 
external physical fields that couple directly to ${\bm{\kappa}}_{jk}$): the chiral components in  multiferroic MnWO$_4$ have been detected by neutron diffraction using 
spherical polarization analysis as a function of temperature and of external electric field \cite{Finger2010}.
The vector chirality, which is the argument of this letter, always accompanies helical magnetic 
order, and it can arise from spontaneous $\mathbb{Z}_2$ symmetry breaking in systems with competitive 
exchange interactions \cite{Kawamura1998}, or it can be stabilized by the Dzyaloshinskii-Moriya antisymmetric exchange 
interaction in noncentrosymmetric compounds, 
\cite{Sergienko2006,Pfleiderer2001,Pedrazzini2007} . However, the vector chiral symmetry can be broken also in a magnetically disordered 
state. Such phase is named a vector chiral spin liquid phase and has been intensively 
studied in the last years. It has been predicted to occur in one-dimensional (1\textit{d}) 
frustrated quantum magnetic 
systems \cite{Hikihara2008,Sudan2009,Furukawa2010}. For higher dimension $d$ it is crucial to understand if the  vector chiral spin 
liquid phase is stable also in presence of thermal fluctuations \cite{Furukawa2010}. For \textit{d}=2, this phase has been 
clearly obtained  at finite temperature $T$ by classical Monte Carlo (MC) simulation of a
triangular lattice of spins with bilinear and biquadratic interactions \cite{Nematic2008}. However, 
for models without geometrical frustration and \textit{d}=2,\,3, a clear evidence of 
this exotic phase is yet lacking, even if Onoda and Nagaosa \cite{Onoda2007,NagaosaReview2008}, investigating a 
Ginzburg-Landau Hamiltonian describing helical magnets, suggest that a vector chiral spin liquid 
phase can be stabilized  even in \textit{d}=3. This prediction was questioned by  Okubo and Kawamura 
\cite{Okubo2010}, because the results of their classical MC simulations do not show any evidence of 
such phase, but only a first order phase transition to a helimagnetic order. In this context, it is 
important to note that in the quasi-1\textit{d} $XY$ organic magnet Gd(hfac)$_3$NITEt \cite{GdEt} (a compound 
with high value of spin, \textit{S}=7/2) a 3\textit{d} vector chiral spin liquid phase has been experimentally 
observed. This result is due to the fact that \textit{d}=1 is the lower critical dimension for an Ising 
order parameter, like  $\langle{{\bm 
\kappa}_{jk}}\rangle$, so that the chiral correlation length, which diverges 
exponentially at low \textit{T}, is much larger than the spin correlation length, which diverges with a 
power law. Taking into account  the interchain interaction  
within mean field approximation a $3d$ vector chiral spin-liquid phase results at intermediate \textit{T} 
\cite{Villain1978}. However, theoretical results obtained considering also the interchain 
fluctuations are still lacking, and a direct numerical evidence of such a chiral phase in quasi-1\textit{d} 
system will be relevant.
\\ \indent
In this letter we present the results obtained by employing accurate classical
MC simulation techniques to investigate a $2d$ spin system composed by weakly 
interacting helimagnetic chains.
Despite the absence of geometrical frustration (the model being defined on a square lattice), 
a clear separation can be observed between the chiral phase transition temperature $T_\kappa$ and 
the Kosterlitz-Thouless (\textit{KT}) one $T_{KT}$ separating the quasi-long-range spin ordered 
phase from the disordered one.
\\ \indent
We consider a simple square lattice on the $(x,y)$-plane composed of 
$N$=$L$$\times$$L$ planar spins $\vec{S}_{i,j}$  
($|\vec{S}|$=1), whose interactions are described by the Hamiltonian:
\begin{equation}
\label{ham}
\begin{split} 
{\cal H}=-\sum_{i=1}^{L}\sum_{j=1}^{L} &
\left\{J_1\vec{S}_{i,j}\cdot\vec{S}_{i,j+1} + J_2\vec{S}_{i,j}\cdot\vec{S}_{i,j+2}\right.\\
&\left.+J^\prime \vec{S}_{i,j}\cdot\vec{S}_{i+1,j}\right\}\,;
\end{split}
\end{equation}
\textit{j} labels spins along each chain, while \textit{i} is the chain label.
Intra-chain exchange interactions are ruled by a nearest neighbour (NN), ferromagnetic (FM) 
coupling constant $J_1$ and a next-nearest neighbour (NNN), antiferromagnetic (AFM)
coupling $J_2$; interchain NN spin interactions are ruled by the FM coupling constant 
$J^\prime$. If the condition $\delta$$\equiv$$|J_2|$/$J_1$$>$1/4 
is fulfilled, the ground state corresponds to a 
helical order along the chains, with a pitch vector
$q_{\parallel}$=$\pm\cos^{-1}\left(-1/4\delta\right)$ . 
In the following we take $\delta=0.3$ (i.e. $J_1$=1, and $|J_2|$=0.3),
while $T$ will be given in units $J_1$. 
Periodic boundary conditions have been applied along the direction 
perpendicular to the chains while, due to the incommensurate helix 
modulation, free boundary conditions were taken along the chain 
direction.
Configuration sampling has been carried on making use of the usual
Metropolis technique, while correlations between 
sampled configurations were mitigated by microcanonical 
over-relaxed moves\cite{Landau2005B}.
For each $T$ at least three different runs have been performed, each
run being composed of $24\times 10^{6}$ MC sweeps,
with the first $4\times 10^6$ thermalization steps being discarded.
Near the critical regions the multiple-histogram (MH) 
methods \cite{Landau2005B} were employed.
Results for $L=24-128$ are reported.
\begin{figure} [t]
\begin{center}              
\includegraphics[width=9cm]{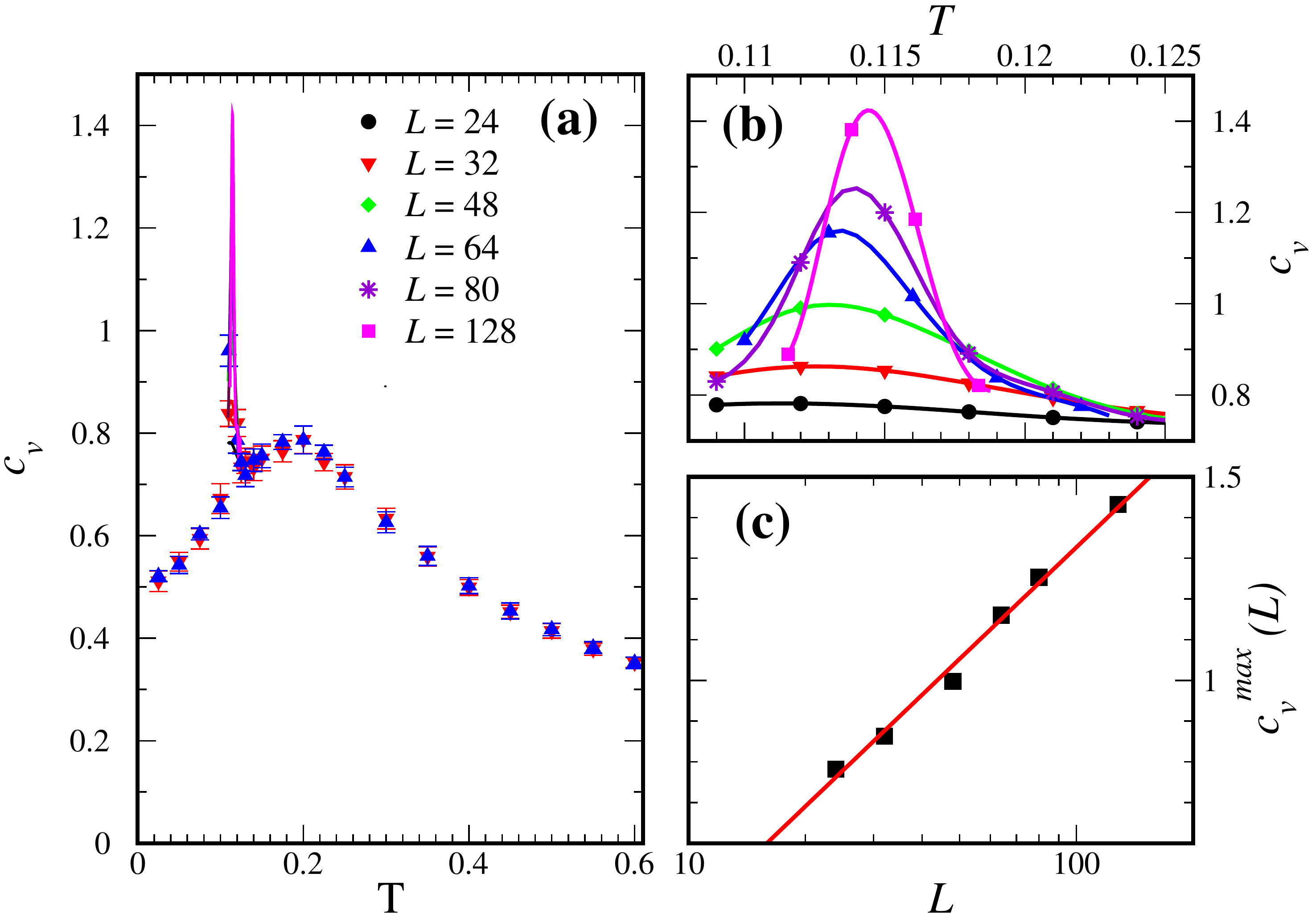}
\end{center}
\caption{(Color online) (a) Specific heat $vs.$ $T$ for the model 
\eqref{ham} with $J^\prime$=0.1$J$ for different $L$ (see legend). 
(b) MH result around $T_\kappa$ for different $L$.
(c) Maximum of the specific heat $vs.$ $L$ 
(error bars lie within point size). Continuum line: fit function, see text.}
\label{fig1}
\end{figure}
\\ \indent
The weakly interacting chain system has been investigated by setting $J^\prime$=0.1$J$: 
In Fig.~\ref{fig1}a the obtained results for the specific heat, $c_v$, \textit{vs.} $T$ are reported. We
observe a well defined narrow and sharp peak at $T\simeq 0.12$, which should 
be ascribed to the onset of a vector chiral spin liquid phase.
In this $T$ range, the scaling behavior of $c_v$ for different 
$L$ is reported in Fig.~\ref{fig1}b: 
We immediately observe as increasing $L$ 
the peak more and more acquires the typical features associated 
with a second order phase transition in the thermodynamic limit. 
Increasing $T$, 
a second broad and size-independent peak is observed at $T\simeq 0.2$,
which is consistent with a \textit{KT} scenario.
\begin{figure} [t]
\begin{center}              
\includegraphics[width=8cm]{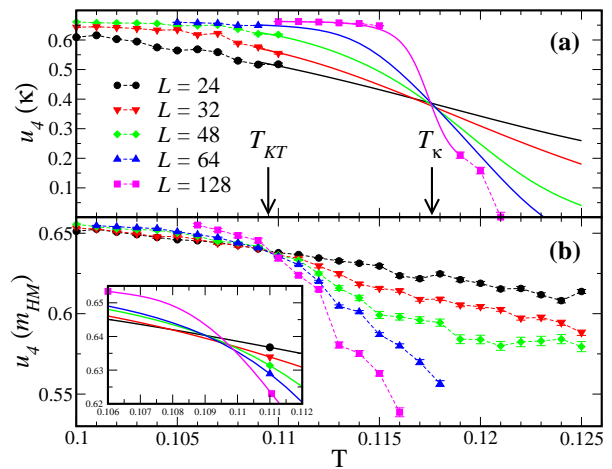}
\end{center}
\caption{(Color online) (a) Binder cumulant for the order parameter 
$\kappa$ $vs.$ $T$ for different lattice sizes; continuous lines: 
MH interpolation. (b) Binder cumulant for the helical order 
parameter $vs.$ temperature for different lattice sizes. Inset:
MH interpolation around $T_{KT}$. }
\label{fig2}
\end{figure} 
In order to estimate the critical temperatures,
we employ the Binder's fourth cumulant 
$u_4$=1$-$$\langle {\cal O}^4 \rangle/3\langle {\cal O}^2 \rangle $ \cite{Landau2005B},
where ${\cal O}$ will be the chirality, $\kappa$=$K^{\prime}\sum_{ij}\left[\vec{S}_{ij}\times\vec{S}_{ij+1}\right]^z$
(where $K^{\prime}$=$\left[L(L-1)\sin q_{\parallel}\right]^{-1}$), or 
the helical order parameter, $m_{HM}$, defined as
$m_{HM}$=$K^{\prime\prime}$$\int dq_{ \parallel}S(\vec{q})$,
where $S(\vec{q})$ is the structure factor, with $\vec{q}$=$(0,q_{\parallel})$, 
and the normalization factor $K^{\prime\prime}$ is the reciprocal
of the structure factor integral at $T$=0 \cite{CRC10}.
The Binder cumulant for different $L$ is reported in 
Fig.~\ref{fig2}a and b for the chirality 
and the helical order parameters, respectively. 
For the chirality we can evaluate $T_\kappa$=$0.1176(6)$,
while for the helical order parameter,   
we obtain $T_{KT}$=$0.1095(5)$. 
The data in Fig.~\ref{fig2} allows us to assert that the two critical temperatures are 
well distinguishable with $(T_\kappa - T_{KT})/T_{KT}\simeq 7.4\%$.
\begin{figure} [t]
\begin{center}              
\includegraphics[width=8cm]{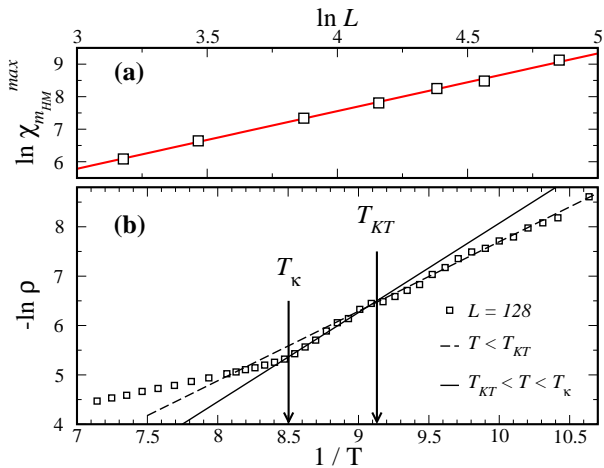}
\end{center}
\caption{(Color online) (a) Logarithm of the maximum of $\chi_{m_{HM}}$ 
as a function of $\ln L$. The error bars lie within the symbols. 
(b) Vortex density $\rho$ $vs.$ $T^{-1}$ for $L$=128, error bars lie within 
point size. Continuum line: linear regression in the chiral region. 
Dashed line: linear regression in the \textit{KT} regime.}
\label{fig3}
\end{figure} 
The identification of the crossing temperature in Fig.~\ref{fig2}b
as the \textit{KT} transition temperature can be further validated by 
making use of the finite-size scaling (FSS) relation  
$\chi_{m}(L)\propto L^{\gamma/\nu}$; a best fit procedure gives 
$\gamma/\nu$=$1.77\,(3)$ (Fig.~\ref{fig3}a), fully consistent 
with the \textit{KT} behaviour of a $2d$ planar system.
Another evidence of the presence of two 
distinct critical points comes from the scrutiny of the vortex density, $\rho$ \footnote{Free boundary conditions along the chain direction can imply the existence of unpaired vortex (antivortex). However, their contribution to the vortex density results absolutely negligible.}.
In the dilute-gas approximation, we have 
$\rho$$\sim$$\exp(-2\mu /T)$, where $2\mu$ is the energy required to create a
pair of vortices \cite{Nelson1983}, and it can be obtained by linear fit 
of $-\ln \rho$ as a function of $T^{-1}$ (Fig.~\ref{fig3}b).
Three different regimes can be identified: low-$T$ ($T$$<$$T_{KT}$), 
intermediate-$T$ ($T_{KT}$$<$$T$$<$$T_{\kappa}$), 
and high-$T$ regime ($T_\kappa$$<$$T$).
The linear fit in the range $T_{KT}$$<$$T$$<$$T_{\kappa}$ (solid line)
shows an activation energy of dissociated vortex pairs greater than that obtained for lower 
temperatures $T$$<$$T_{KT}$, where all vortex pairs are bounded,
with a clear slope change at $T$$\simeq$$T_{KT}$.
Finally, the creation of other dissociated vortex pairs appears again easier 
in the region $T$$>$$T_{\kappa}$,
where $\mu$ strongly decreases signaling the onset of a complete disorder \cite{Gupta1992}.
\\ \indent
A proper characterization of the vector chiral spin liquid transition involves several 
aspects. A first issue, concerning the order of the transition,
can be coped with by analyzing the equilibrium energy distribution 
at $T$=$T_\kappa$. 
Even at the largest simulated size of the lattice, $L$=128, 
no double-peaked structure was observed, so that we have no explicit 
indication in favor of classifying the chiral transition as
a first-order one.
\\ \indent
\begin{figure} [t]
\begin{center}              
\includegraphics[width=8cm]{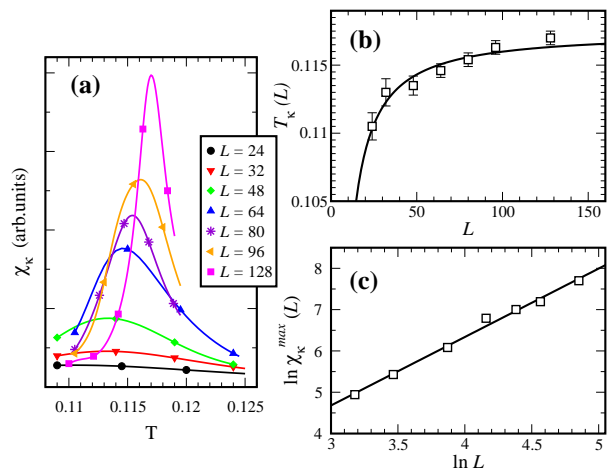}
\end{center}
\caption{(Color online) 
(a) Chiral susceptibility $vs.$ $T$ by MH interpolation for different sizes.   
In (b) the temperature where the peak of $\chi_\kappa$ is located is reported  
$vs.$ $L$.
In (c) the maximum value of $\chi_\kappa$ is reported as a function 
of $\ln L$. Error bars fall within the symbols.}
\label{fig4}
\end{figure} 
The universality class pertaining to the vector chiral spin liquid transition 
has been investigated by an accurate 
FSS analysis. 
In Fig.~\ref{fig4}a the chiral susceptibility, $\chi_\kappa$, is displayed for different values of $L$.
From the expected dependence of the peak position temperature on $L$,  
$T_\kappa(L)$=$T_\kappa+cL^{-1/\nu}$, we can estimate the critical exponent $\nu$,
making use of the value of $T_\kappa$ previously obtained from the Binder's 
cumulant discussed above, getting $\nu$=$1.02(5)$ (Fig.~\ref{fig4}b). 
Analyzing the peak values of $\chi_\kappa$
with the FSS relation $\chi_\kappa(L)$$\propto$$ L^{\gamma/\nu}$,
we obtain the ratio $\gamma/\nu$=1.66(7) 
(Fig.~\ref{fig4}c), which implies $\gamma$=1.70(8). 
These values of $\gamma$ 
and $\nu$ are in very fair agreement 
with $\gamma$=7/4 and $\nu$=1, i.e. the proper values of the Ising 
universality class in $2d$. Concerning the critical exponent $\alpha$ for $c_v$, 
which for the Ising universality class in $2d$ is $0$, the 
$c_v$-peak values, $c^{max}_v(L)$, $vs.$ $L$ are very well fitted (see Fig. 1c) by
the FSS relation proper of the $2d$ Ising model 
$c^{max}_v(L)$=$A$+$B\ln(L)$+$CL^{-1}$,\cite{Barber1983}. This, allows us to conclude 
that $\alpha$=0, confirming the Ising character of 
the vector chiral spin liquid transition.
\\ \indent
We point out that the quasi-$1d$ nature of the model is fundamental in order 
to obtain a vector chiral spin liquid phase in absence of (quasi-)long-range order. 
This has been explicitly checked by MC simulations we have performed assuming 
$J^\prime$=$J_1$, a model investigated by Garel and Doniach \cite{Garel1980} many years ago. In Fig.~\ref{fig0}a-d a summary of the obtained results is reported.
\begin{figure} [t]
\begin{center}              
\includegraphics[width=9cm]{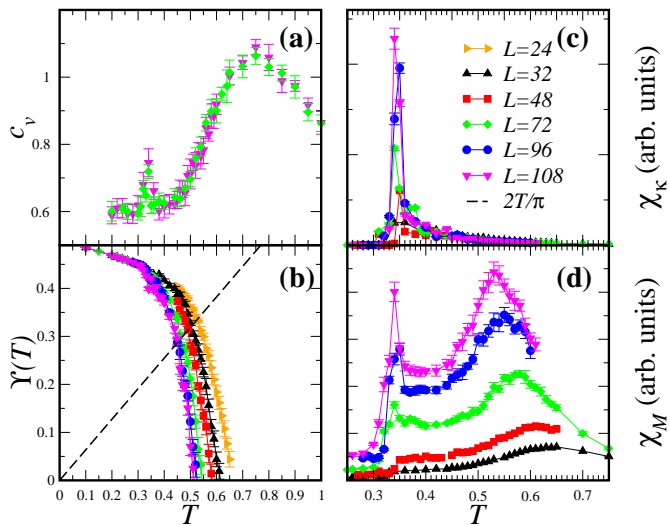}
\end{center}
\caption{(Color online) Observables calculated for the system with $J_1$=$J^\prime$ 
for different size $L$: (a) specific heat; (b) helicity modulus; (c) chiral susceptibility; (d) 
susceptibility of $M$ (see text). }
\label{fig0}
\end{figure}
\\ \indent
For the specific heat, $c_v$, reported 
\textit{vs.} $T$, we observe a size-independent broad peak
at $T$$\simeq$0.75, consistent with a \textit{KT} scenario, while a 
size-dependent, narrow peak at $T$$\simeq$0.34 is found.
Using the helicity modulus $\Upsilon(T)$ \cite{chaikin2000principles},
one is able to evaluate  $T_{KT}$ taking advantage of the
universal jump:
$\Upsilon(T)/T\to 2/ \pi$ for $T\to T_{KT}$.
We estimate $T_{KT}$$\simeq$0.45. 
For the largest simulated size ($L$=108), 
$\chi_\kappa$ shows 
a narrow peak at the same temperature $T_\kappa$ of the narrow,
size-dependent, $c_v$ peak and, above all, $T_\kappa$ is significantly lower 
than $T_{KT}$.
These observations are corroborated by the data obtained 
for a new parameter defined as $M$=$\frac{1}{L}\sum_{i=1}^L m_i$
where $m_i$=$\sqrt{(\frac{1}{L}\sum_{j=1}^L S^x_{i,j})^2+
(\frac{1}{L}\sum_{j=1}^L S^y_{i,j})^2}$ 
is the columnar magnetization perpendicular to the helical displacement.
This observable turns out to be relevant both for the chiral 
phase and for the  establishment of the $KT$ phase:
Indeed, both two- and four-points correlations contribute to its susceptibility 
$\chi_M$ (Fig.~\ref{fig0}d), which displays 
a first anomaly, which progressively stabilizes at $T_{KT}$ when $L$ increases,
signaling the onset of a quasi-order;
subsequently, at lower-$T$, $\chi_M$ has a second anomaly consistent
with those already displayed by $c_v$ and $\chi_\kappa$. So, we can definitely 
estimate $T_\kappa$$\simeq$0.34.
We conclude that, for $J^\prime$=$J_1$, we have
a clear separation between the \textit{KT} behaviour and chiral setup, but, at variance with the quasi-$1d$ case,
the onset of the chiral order is established at a temperature 
$T_\kappa$ lower than $T_{KT}$.
\\ \indent
In conclusion, we have presented the outcomes of   
intensive MC simulations  for a $2d$ \textit{XY} classical spin system, defined on a square lattice, composed by 
weakly interacting frustrated chains. We observe a clear separation between the vector chiral spin liquid phase and the quasi-long-range ordered phase, with $T_\kappa$$>$$T_{KT}$.
We have found that, in a system without geometrical frustration, the chirality  
displays a second order phase transition, consistent with the $2d$ Ising universality class. 
This result confirms the intriguing possibility of an emergent finite-temperature phase showing 
chiral long range order in the absence of the helical one as investigated by many authors in the 
multiferroic context \cite{Onoda2007,Nematic2008,Furukawa2010,NagaosaReview2008}.
We found the quasi-$1d$ nature of the 
system being fundamental in order to observe such an exotic phase in absence of (quasi-)long-range order: indeed, assuming the same NN exchange 
constants in both directions ($J^\prime$=$J_1$) we find that  the sequence of phase transitions can be reversed  $T_\kappa$$<$$T_{KT}$.
The opposite sequence of the two phase transitions for the investigated $2d$ models 
can also give indications for the behavior of their $3d$ counterparts. As we move
from $d$=2 to $d$=3, the $KT$ phase transition is replaced by a proper second order 
phase transition to a helimagnetic spin arrangement which implies an underlying 
chiral one, so that it is not surprising that in their MC simulation 
Okubo and Kawamura \cite{Okubo2010} do not observe a chiral phase but only the 
phase transition to the helical order, which also entails chiral order. On the contrary, 
for a $3d$ collection of 
weakly interacting helimagnetic chains, it is reasonable to hypothesize that 
the phase transition to helimagnetic order occurs at a lower $T$ than the chiral 
one, and the vector chiral spin-liquid phase can manifest \cite{Villain1978} according to the scenario 
recently emerged from experiments on the quasi-$1d$ organic high-spin magnet Gd(hfac)$_3$NITEt \cite{GdEt}. 
%
%
\\ \indent
Acknowledgments: F.C. thanks INSTM,
the Natural Science and Engineering Research Council of Canada under 
Research Grant 121210893, and the Alberta Informatics Circle of 
Research Excellence (iCore);
financial support 
was also given by the italian Ministry of University within the 2008 PRIN program 
(contract N. 2008PARRTS\_003). 

\bibliography{spin}

\begin{thebibliography}{27}
\expandafter\ifx\csname natexlab\endcsname\relax\def\natexlab#1{#1}\fi
\expandafter\ifx\csname bibnamefont\endcsname\relax
  \def\bibnamefont#1{#1}\fi
\expandafter\ifx\csname bibfnamefont\endcsname\relax
  \def\bibfnamefont#1{#1}\fi
\expandafter\ifx\csname citenamefont\endcsname\relax
  \def\citenamefont#1{#1}\fi
\expandafter\ifx\csname url\endcsname\relax
  \def\url#1{\texttt{#1}}\fi
\expandafter\ifx\csname urlprefix\endcsname\relax\def\urlprefix{URL }\fi
\providecommand{\bibinfo}[2]{#2}
\providecommand{\eprint}[2][]{\url{#2}}

\bibitem[{\citenamefont{Machida et~al.}(2007)\citenamefont{Machida, Nakatsuji,
  Maeno, Tayama, Sakakibara, and Onoda}}]{Machida2007}
\bibinfo{author}{\bibfnamefont{Y.}~\bibnamefont{Machida}},
  \bibinfo{author}{\bibfnamefont{S.}~\bibnamefont{Nakatsuji}},
  \bibinfo{author}{\bibfnamefont{Y.}~\bibnamefont{Maeno}},
  \bibinfo{author}{\bibfnamefont{T.}~\bibnamefont{Tayama}},
  \bibinfo{author}{\bibfnamefont{T.}~\bibnamefont{Sakakibara}},
  \bibnamefont{and} \bibinfo{author}{\bibfnamefont{S.}~\bibnamefont{Onoda}},
  \bibinfo{journal}{Phys. Rev. Lett.} \textbf{\bibinfo{volume}{98}},
  \bibinfo{pages}{057203} (\bibinfo{year}{2007}).

\bibitem[{\citenamefont{Bulaevskii et~al.}(2008)\citenamefont{Bulaevskii,
  Batista, Mostovoy, and Khomskii}}]{Bulaevskii2008}
\bibinfo{author}{\bibfnamefont{L.~N.} \bibnamefont{Bulaevskii}},
  \bibinfo{author}{\bibfnamefont{C.~D.} \bibnamefont{Batista}},
  \bibinfo{author}{\bibfnamefont{M.~V.} \bibnamefont{Mostovoy}},
  \bibnamefont{and} \bibinfo{author}{\bibfnamefont{D.~I.}
  \bibnamefont{Khomskii}}, \bibinfo{journal}{Phys. Rev. B}
  \textbf{\bibinfo{volume}{78}}, \bibinfo{pages}{024402}
  (\bibinfo{year}{2008}).

\bibitem[{\citenamefont{Al-Hassanieh et~al.}(2009)\citenamefont{Al-Hassanieh,
  Batista, Ortiz, and Bulaevskii}}]{Al-Hassanieh2009}
\bibinfo{author}{\bibfnamefont{K.~A.} \bibnamefont{Al-Hassanieh}},
  \bibinfo{author}{\bibfnamefont{C.~D.} \bibnamefont{Batista}},
  \bibinfo{author}{\bibfnamefont{G.}~\bibnamefont{Ortiz}}, \bibnamefont{and}
  \bibinfo{author}{\bibfnamefont{L.~N.} \bibnamefont{Bulaevskii}},
  \bibinfo{journal}{Phys. Rev. Lett.} \textbf{\bibinfo{volume}{103}},
  \bibinfo{pages}{216402} (\bibinfo{year}{2009}).

\bibitem[{\citenamefont{Katsura et~al.}(2005)\citenamefont{Katsura, Nagaosa,
  and Balatsky}}]{Katsura2005}
\bibinfo{author}{\bibfnamefont{H.}~\bibnamefont{Katsura}},
  \bibinfo{author}{\bibfnamefont{N.}~\bibnamefont{Nagaosa}}, \bibnamefont{and}
  \bibinfo{author}{\bibfnamefont{A.~V.} \bibnamefont{Balatsky}},
  \bibinfo{journal}{Phys. Rev. Lett.} \textbf{\bibinfo{volume}{95}},
  \bibinfo{pages}{057205} (\bibinfo{year}{2005}).

\bibitem[{\citenamefont{Mostovoy}(2006)}]{Mostovoy2006}
\bibinfo{author}{\bibfnamefont{M.}~\bibnamefont{Mostovoy}},
  \bibinfo{journal}{Phys. Rev. Lett.} \textbf{\bibinfo{volume}{96}},
  \bibinfo{eid}{067601} (\bibinfo{year}{2006}).

\bibitem[{\citenamefont{Sergienko and Dagotto}(2006)}]{Sergienko2006}
\bibinfo{author}{\bibfnamefont{I.~A.} \bibnamefont{Sergienko}}
  \bibnamefont{and} \bibinfo{author}{\bibfnamefont{E.}~\bibnamefont{Dagotto}},
  \bibinfo{journal}{Phys. Rev. B} \textbf{\bibinfo{volume}{73}},
  \bibinfo{pages}{094434} (\bibinfo{year}{2006}).

\bibitem[{\citenamefont{Cheong and Mostovoy}(2007)}]{Cheong2007}
\bibinfo{author}{\bibfnamefont{S.}~\bibnamefont{Cheong}} \bibnamefont{and}
  \bibinfo{author}{\bibfnamefont{M.}~\bibnamefont{Mostovoy}},
  \bibinfo{journal}{Nature materials} \textbf{\bibinfo{volume}{6}},
  \bibinfo{pages}{13} (\bibinfo{year}{2007}).

\bibitem[{\citenamefont{Finger et~al.}(2010)\citenamefont{Finger, Senff,
  Schmalzl, Schmidt, Regnault, Becker, Bohaty, and Braden}}]{Finger2010}
\bibinfo{author}{\bibfnamefont{T.}~\bibnamefont{Finger}},
  \bibinfo{author}{\bibfnamefont{D.}~\bibnamefont{Senff}},
  \bibinfo{author}{\bibfnamefont{K.}~\bibnamefont{Schmalzl}},
  \bibinfo{author}{\bibfnamefont{W.}~\bibnamefont{Schmidt}},
  \bibinfo{author}{\bibfnamefont{L.}~\bibnamefont{Regnault}},
  \bibinfo{author}{\bibfnamefont{P.}~\bibnamefont{Becker}},
  \bibinfo{author}{\bibfnamefont{L.}~\bibnamefont{Bohaty}}, \bibnamefont{and}
  \bibinfo{author}{\bibfnamefont{M.}~\bibnamefont{Braden}},
  \bibinfo{journal}{J. Phys.: Conf. Series} \textbf{\bibinfo{volume}{211}},
  \bibinfo{pages}{012001} (\bibinfo{year}{2010}).

\bibitem[{\citenamefont{Kawamura}(1998)}]{Kawamura1998}
\bibinfo{author}{\bibfnamefont{H.}~\bibnamefont{Kawamura}},
  \bibinfo{journal}{J. Phys.: Cond. Matt.} \textbf{\bibinfo{volume}{10}},
  \bibinfo{pages}{4707} (\bibinfo{year}{1998}).

\bibitem[{\citenamefont{Pfleiderer et~al.}(2001)\citenamefont{Pfleiderer,
  Julian, and Lonzarich}}]{Pfleiderer2001}
\bibinfo{author}{\bibfnamefont{C.}~\bibnamefont{Pfleiderer}},
  \bibinfo{author}{\bibfnamefont{S.~R.} \bibnamefont{Julian}},
  \bibnamefont{and} \bibinfo{author}{\bibfnamefont{G.~G.}
  \bibnamefont{Lonzarich}}, \bibinfo{journal}{Nature}
  \textbf{\bibinfo{volume}{414}}, \bibinfo{pages}{427} (\bibinfo{year}{2001}).

\bibitem[{\citenamefont{Pedrazzini et~al.}(2007)\citenamefont{Pedrazzini,
  Wilhelm, Jaccard, Jarlborg, Schmidt, Hanfland, Akselrud, Yuan, Schwarz, Grin
  et~al.}}]{Pedrazzini2007}
\bibinfo{author}{\bibfnamefont{P.}~\bibnamefont{Pedrazzini}},
  \bibinfo{author}{\bibfnamefont{H.}~\bibnamefont{Wilhelm}},
  \bibinfo{author}{\bibfnamefont{D.}~\bibnamefont{Jaccard}},
  \bibinfo{author}{\bibfnamefont{T.}~\bibnamefont{Jarlborg}},
  \bibinfo{author}{\bibfnamefont{M.}~\bibnamefont{Schmidt}},
  \bibinfo{author}{\bibfnamefont{M.}~\bibnamefont{Hanfland}},
  \bibinfo{author}{\bibfnamefont{L.}~\bibnamefont{Akselrud}},
  \bibinfo{author}{\bibfnamefont{H.~Q.} \bibnamefont{Yuan}},
  \bibinfo{author}{\bibfnamefont{U.}~\bibnamefont{Schwarz}},
  \bibinfo{author}{\bibfnamefont{Y.}~\bibnamefont{Grin}}, \bibnamefont{et~al.},
  \bibinfo{journal}{Phys. Rev. Lett.} \textbf{\bibinfo{volume}{98}},
  \bibinfo{pages}{047204} (\bibinfo{year}{2007}).

\bibitem[{\citenamefont{Hikihara et~al.}(2008)\citenamefont{Hikihara, Kecke,
  Momoi, and Furusaki}}]{Hikihara2008}
\bibinfo{author}{\bibfnamefont{T.}~\bibnamefont{Hikihara}},
  \bibinfo{author}{\bibfnamefont{L.}~\bibnamefont{Kecke}},
  \bibinfo{author}{\bibfnamefont{T.}~\bibnamefont{Momoi}}, \bibnamefont{and}
  \bibinfo{author}{\bibfnamefont{A.}~\bibnamefont{Furusaki}},
  \bibinfo{journal}{Phys. Rev. B} \textbf{\bibinfo{volume}{78}},
  \bibinfo{pages}{144404} (\bibinfo{year}{2008}).

\bibitem[{\citenamefont{Sudan et~al.}(2009)\citenamefont{Sudan, L\"uscher, and
  L\"auchli}}]{Sudan2009}
\bibinfo{author}{\bibfnamefont{J.}~\bibnamefont{Sudan}},
  \bibinfo{author}{\bibfnamefont{A.}~\bibnamefont{L\"uscher}},
  \bibnamefont{and} \bibinfo{author}{\bibfnamefont{A.~M.}
  \bibnamefont{L\"auchli}}, \bibinfo{journal}{Phys. Rev. B}
  \textbf{\bibinfo{volume}{80}}, \bibinfo{pages}{140402}
  (\bibinfo{year}{2009}).

\bibitem[{\citenamefont{Furukawa et~al.}(2010)\citenamefont{Furukawa, Sato, and
  Onoda}}]{Furukawa2010}
\bibinfo{author}{\bibfnamefont{S.}~\bibnamefont{Furukawa}},
  \bibinfo{author}{\bibfnamefont{M.}~\bibnamefont{Sato}}, \bibnamefont{and}
  \bibinfo{author}{\bibfnamefont{S.}~\bibnamefont{Onoda}},
  \bibinfo{journal}{arXiv:1003,3940v1}  (\bibinfo{year}{2010}).

\bibitem[{\citenamefont{Park et~al.}(2008)\citenamefont{Park, Onoda, Nagaosa,
  and Han}}]{Nematic2008}
\bibinfo{author}{\bibfnamefont{J.-H.} \bibnamefont{Park}},
  \bibinfo{author}{\bibfnamefont{S.}~\bibnamefont{Onoda}},
  \bibinfo{author}{\bibfnamefont{N.}~\bibnamefont{Nagaosa}}, \bibnamefont{and}
  \bibinfo{author}{\bibfnamefont{J.~H.} \bibnamefont{Han}},
  \bibinfo{journal}{Phys. Rev. Lett.} \textbf{\bibinfo{volume}{101}},
  \bibinfo{eid}{167202} (\bibinfo{year}{2008}).

\bibitem[{\citenamefont{Onoda and Nagaosa}(2007)}]{Onoda2007}
\bibinfo{author}{\bibfnamefont{S.}~\bibnamefont{Onoda}} \bibnamefont{and}
  \bibinfo{author}{\bibfnamefont{N.}~\bibnamefont{Nagaosa}},
  \bibinfo{journal}{Phys. Rev. Lett.} \textbf{\bibinfo{volume}{99}},
  \bibinfo{eid}{027206} (\bibinfo{year}{2007}).

\bibitem[{\citenamefont{Nagaosa}(2008)}]{NagaosaReview2008}
\bibinfo{author}{\bibfnamefont{N.}~\bibnamefont{Nagaosa}}, \bibinfo{journal}{J.
  Phys.: Cond. Matt.} \textbf{\bibinfo{volume}{20}}, \bibinfo{pages}{434207}
  (\bibinfo{year}{2008}).

\bibitem[{\citenamefont{Okubo and Kawamura}(2010)}]{Okubo2010}
\bibinfo{author}{\bibfnamefont{T.}~\bibnamefont{Okubo}} \bibnamefont{and}
  \bibinfo{author}{\bibfnamefont{H.}~\bibnamefont{Kawamura}},
  \bibinfo{journal}{arXiv:1001.4662v1}  (\bibinfo{year}{2010}).

\bibitem[{\citenamefont{Cinti et~al.}(2008)\citenamefont{Cinti, Rettori, Pini,
  Mariani, Micotti, Lascialfari, Papinutto, Amato, Caneschi, Gatteschi
  et~al.}}]{GdEt}
\bibinfo{author}{\bibfnamefont{F.}~\bibnamefont{Cinti}},
  \bibinfo{author}{\bibfnamefont{A.}~\bibnamefont{Rettori}},
  \bibinfo{author}{\bibfnamefont{M.~G.} \bibnamefont{Pini}},
  \bibinfo{author}{\bibfnamefont{M.}~\bibnamefont{Mariani}},
  \bibinfo{author}{\bibfnamefont{E.}~\bibnamefont{Micotti}},
  \bibinfo{author}{\bibfnamefont{A.}~\bibnamefont{Lascialfari}},
  \bibinfo{author}{\bibfnamefont{N.}~\bibnamefont{Papinutto}},
  \bibinfo{author}{\bibfnamefont{A.}~\bibnamefont{Amato}},
  \bibinfo{author}{\bibfnamefont{A.}~\bibnamefont{Caneschi}},
  \bibinfo{author}{\bibfnamefont{D.}~\bibnamefont{Gatteschi}},
  \bibnamefont{et~al.}, \bibinfo{journal}{Phys. Rev. Lett.}
  \textbf{\bibinfo{volume}{100}}, \bibinfo{pages}{057203}
  (\bibinfo{year}{2008}).

\bibitem[{\citenamefont{Villain}(1978)}]{Villain1978}
\bibinfo{author}{\bibfnamefont{J.}~\bibnamefont{Villain}},
  \bibinfo{journal}{Ann. Isr. Phys. Soc.} \textbf{\bibinfo{volume}{2}},
  \bibinfo{pages}{565} (\bibinfo{year}{1978}).

\bibitem[{\citenamefont{Landau and Binder}(2005)}]{Landau2005B}
\bibinfo{author}{\bibfnamefont{D.}~\bibnamefont{Landau}} \bibnamefont{and}
  \bibinfo{author}{\bibfnamefont{K.}~\bibnamefont{Binder}},
  \emph{\bibinfo{title}{{A guide to Monte Carlo simulations in statistical
  physics}}} (\bibinfo{publisher}{Cambridge Univ Pr}, \bibinfo{year}{2005}).

\bibitem[{\citenamefont{Cinti et~al.}(2010)\citenamefont{Cinti, Rettori, and
  Cuccoli}}]{CRC10}
\bibinfo{author}{\bibfnamefont{F.}~\bibnamefont{Cinti}},
  \bibinfo{author}{\bibfnamefont{A.}~\bibnamefont{Rettori}}, \bibnamefont{and}
  \bibinfo{author}{\bibfnamefont{A.}~\bibnamefont{Cuccoli}},
  \bibinfo{journal}{Phys. Rev. B} \textbf{\bibinfo{volume}{81}},
  \bibinfo{pages}{134415} (\bibinfo{year}{2010}).

\bibitem[{\citenamefont{Nelson}(1983)}]{Nelson1983}
\bibinfo{author}{\bibfnamefont{D.~R.} \bibnamefont{Nelson}},
  vol.~\bibinfo{volume}{7} of \emph{\bibinfo{series}{Phase Transition and
  Critical Phenomena}} (\bibinfo{publisher}{Academic Press},
  \bibinfo{address}{New York}, \bibinfo{year}{1983}), \bibinfo{note}{and
  references therein.}

\bibitem[{\citenamefont{Gupta and Baillie}(1992)}]{Gupta1992}
\bibinfo{author}{\bibfnamefont{R.}~\bibnamefont{Gupta}} \bibnamefont{and}
  \bibinfo{author}{\bibfnamefont{C.~F.} \bibnamefont{Baillie}},
  \bibinfo{journal}{Phys. Rev. B} \textbf{\bibinfo{volume}{45}},
  \bibinfo{pages}{2883} (\bibinfo{year}{1992}).

\bibitem[{\citenamefont{Barber}(1983)}]{Barber1983}
\bibinfo{author}{\bibfnamefont{M.~N.} \bibnamefont{Barber}},
  vol.~\bibinfo{volume}{8} of \emph{\bibinfo{series}{Phase Transion and
  Critical Phenomena}} (\bibinfo{publisher}{Academic Press},
  \bibinfo{address}{New York}, \bibinfo{year}{1983}).

\bibitem[{\citenamefont{Garel and Doniach}(1980)}]{Garel1980}
\bibinfo{author}{\bibfnamefont{T.}~\bibnamefont{Garel}} \bibnamefont{and}
  \bibinfo{author}{\bibfnamefont{S.}~\bibnamefont{Doniach}},
  \bibinfo{journal}{J. Phys. C: Sol. St. Phys.} \textbf{\bibinfo{volume}{13}},
  \bibinfo{pages}{L887} (\bibinfo{year}{1980}).

\bibitem[{\citenamefont{Chaikin and Lubensky}(2000)}]{chaikin2000principles}
\bibinfo{author}{\bibfnamefont{P.}~\bibnamefont{Chaikin}} \bibnamefont{and}
  \bibinfo{author}{\bibfnamefont{T.}~\bibnamefont{Lubensky}},
  \emph{\bibinfo{title}{{Principles of condensed matter physics}}}
  (\bibinfo{publisher}{Cambridge Univ Pr}, \bibinfo{year}{2000}).

\end{thebibliography}

\end{document}